\documentclass[11pt,twoside]{article} 
\usepackage{times,fancyhdr}
\usepackage{cite}
\usepackage[english]{babel}

\date{}
\title{Boltzmann and the art of flying} 
\author{S\'{\i}lvio R. Dahmen}
\begin{document} 
\maketitle

\begin{center}
\small{\textit{Instituto de F\'{\i}sica, Universidade Federal do
Rio Grande do Sul, Brazil\\and\\Institut f\"ur Theoretische Physik, Universit\"at W\"urzburg, Germany}}
\end{center}

\thispagestyle{empty}
\begin{abstract}
{\small

One of the less known facets of Ludwig Boltzmann was that of an advocate of Aviation, one of the most
challenging technological problems of his times. Boltzmann followed closely the studies of pioneers like Otto Lilienthal in Berlin, and during a lecture on a prestigious conference he vehemently defended further investments in the area. In this article I discuss his involvement with Aviation, his role in its development and his correspondence with two flight pioneers, Otto Lilienthal e Wilhelm Kress.\\
\textbf{Keywords: History of Physics, Aeronautics} .\\

}

\end{abstract}

\section{Introduction}

A hundred years ago the life of one of the most prolific and influential of XIX century's physicists came to a tragic end. It has long been discussed by historians and biographers the many reasons why Ludwig Eduard Boltzmann decided to put an end to his life while on vacation in Duino~\cite{broda83,flamm73}. After all, notwithstanding the many controversies which raged about his work and the many attacks he had to harness, his career had been a brilliant one: at the age of 25 already a full professor in Graz, by the end of his life he had been appointed to chairs of many distinguished universities, and once even refused changing his family name to the noble `von Boltzmann' with the argument that ``\textit{if our rather middle-class name has been enough for my ancestors, it should remain so for my children and grandchildren}''~\cite{cercignani98}.  While gathering the admiration of many of his contemporaries, some did not fully understand his contributions to Statistical Mechanics, whose foundations he helped establish along with R. Clausius, J.C. Maxwell and particularly J.W. Gibbs, whom he greatly admired~\cite{boltzmannstatmech}. Highly regarded in his native Austria, he travelled extensively overseas. His very spirited account of a trip to the United States also shows a man of keen observational powers and a great sense of humor~\cite{boltzmanneldorado}. 

Here I would like to discuss another facet of Boltzmann, namely that of an advocate of Technology in the
mirror of Aviation. As it happens with many other giants of science, the contributions they are mostly remembered for usually overshadow other fundamental ones. So for instance few remember Heinrich Hertz's Elasticity Theory, a seminal theoretical contribution still extensively employed by practicioners in the field~\cite{hertz}. Isaac Newton's Hydrodynamics pale in comparison to his Mechanics, Gravitation and Optics~\cite{newton}, for reasons I will discuss shortly since they are directly related to the theme of this essay. By the same token, Boltzmann's {\it experimental} contributions to Eletromagnetism have only received the attention they deserved in the last few years~\cite{broda83,cercignani98,rumpf}. These works, albeit important, are seen as secondary to his contributions to Statistical Mechanics. They do however show us a different picture of Boltzmann: that of a first-rate experimental physicist tuned to the technological challenges of his times.

My analysis is based on the lecture ``\"Uber Luftschiffahrt''~\footnote{Literally \textit{On airship travel.} In spite of the word \textit{Luftschiff} having its exact correspondence in the English \textit{Airship}, airplanes are the main theme of his talk.} he delivered during the 66th Annual Meeting of the German Society of Natural Scientists and Physicians held in Vienna in 1894~\cite{boltzmannluft}. Boltzmann was well aware of his stature in the scientific community and seized the opportunity presented by such a privileged venue to passionately defend the brooding field of Aeronautics. This is also confirmed by his correspondence with great aviation pioneers of his time, particularly Otto Lilienthal. I argue that more than representing a passing fancy of a man imbued by the prevailing \textit{Zeitgeist}, the lecture shows Boltzmann's deep concern with the funding of applied research and his attitute towards Technology as an inseparable companion to basic research. 

In the next section I discuss the general context of Boltzmann's early involvement with experimental
problems and the situation of Aeronautics at that time, both experimental and theoretical. Then
I present translations of his correspondence with two aviation pioneers -- Otto Lilienthal and Wilhelm
Kress -- before drawing some conclusions.

\section{A passion for gadgets}

Boltzmann's involvement with technological questions was not uncommon for his times, an age marked by great
technological advances and a positivist atmosphere. In his case though it was also
closely related to the circumstances of his first professorship: while officially appointed to a chair of Mathematical Physics, Boltzmann was actually responsible for the setting up of a General Physics course supposed to meet the needs of an expanding university~\cite{hoeflechner1994}. With the opening of a Medical School in Graz in 1863, the university authorities were faced with the challenge of offering not only more chairs on Natural Sciences but also better ones. Physics was under the responsibility of Karl Hummel, an elder faculty member whose abilities were not up to the new challenges. A decision was made to create a new chair to be occupied by an \textit{Extraordinarius}~\footnote{A professor subordinated to a Full Professor or \textit{Ordinarius}. Nowadays it denotes an associate professor.}. For political reasons this
chair was masked under the name of Mathematical Physics, while it actually foresaw that its
holder would be responsible for equipping and teaching lab courses. This position was shortly occupied until E. Mach, then at Graz, managed
to have it changed to an \textit{Ordinarius} status so that he could move from the Mathematics to
the Physics chair. Shortly after doing this Mach left for Prague in 1867 and Boltzmann replaced him, two years later, in a position he was to remain for the next 4 years until his return
to Vienna~\footnote{He would return to Graz in 1876 and remain there for 14 years. By his own accounts
these were the happiest years of his life~\cite{flamm73}.}. Moreover, he had a professional involvement with experimental research as a consequence of his visit to H. von Helmholtz's Institute in Berlin. Helmholtz, under whose guidance Hertz was introduced to Eletromagnetism, was one of the proponents of Maxwell's Theory in Germany. So it was natural that during his Berlin \textit{sojourn} Helmholtz presented him with problems directly related to the confirmation of Maxwell's Theory. In Boltzmann the \textit{doyen} of Physics in Germany found
fertile ground, for the austrian newcomer had worked on a thesis about the movement of charges on curved surfaces during his PhD and was well acquainted with Maxwell's works, some of which he translated into
German (his supervisor, J. Stefan, was also an advocate of Maxwellian Eletromagnetism)~\cite{boltzmann65}.

Boltzmann's most important contributions to Eletromagnetism took place during the 70's and 80's. His works
fall mainly into two lines: on the one hand he conducted experiments which confirmed Maxwell's theory, thus anticipating Hertz by 14 years ~\cite{boltzmannschwefel}. From a theoretical point of view, he worked on the Hall and Thermoelectric Effects~\cite{boltzmannhall}, Diamagnetism~\cite{boltzmanndiamag} and wrote a book on Maxwell's Electromagnetism~\cite{boltzmannmaxwell}, not to mention the famous Stefan-Boltzmann Law which relates the intensity of the eletromagnetic radiation emitted by a black-body to its temperature~\cite{boltzmann84}. From a total of 31 articles he published on Electromagnetism -- roughly one fifth of his whole scientific production, second only to his output in Statistical Mechanics -- 7 are purely experimental and deal with the relationship between the index of refraction $n$ of a medium
and the dielectric and permeability constants $\epsilon$ and $\mu$ respectively~\cite{dahmen1}.
Maxwell's theory predicted that these quantities were related through:
\begin{equation}
\label{eq:refract}
 n=\sqrt{\epsilon\mu}\;.
\end{equation}
 The relevance of this equation should not be underestimated, for it relates the optical to the electrical
and magnetic properties of a substance, one of the tenets of Maxwellian Eletromagnetism. Starting with sulfur crystals~\cite{boltzmannschwefel} Boltzmann moved to a more challenging task: to measure the
index of refraction and dielectric constant of some gases. Since their indexes of refraction are equal to
$1$ up to the fourth decimal place, one can only confirm Maxwell's prediction by doing high-precision measurements. Boltzmann was up to the task and devised not only new and ingenious ways for measuring these quantities but also the necessary equipment. His results, which in some cases are equal to the values
we use nowadays, can be seen in the following table~\cite{boltzmanngasen}.
\vskip 0.5cm
\begin{center}
\begin{tabular}{l c c}
\hline
Substance&$\sqrt{\epsilon}$& $n$\\ \hline
Air&       1,000295   &1,000294\\
CO$_2$&  1,000473        &1,000449\\
H$_2$&   1,000132         &1,000138\\
CO&      1,000345        &1,000340\\
NO$_2$&  1,000497       &1,000503\\
Oil & 1,000656 &1,000678\\
Swamp gas & 1,000472 &1,000443\\ \hline
\end{tabular} 
\end{center}
For all these substances the permeability $\mu$ is very close to $1$ so that Eq. (\ref{eq:refract}) reduces to $n\sim\sqrt{\epsilon}$. There is no clear indication in his article as to which kind of oil he actually used. As for swamp gas (\textit{Sumpfgas}) he meant Methane (CH$_4$), whose index of refraction
is known to be $1,000443$ (one should keep in mind that $n$ depends on the wavelength and Boltzmann's results are valid for the visible spectrum).

These examples are a clear indication that Boltzmann had actually an insider's knowledge of experimental techniques
and consequently should have been in position to critically assess the experimental data on the growing
but still insipient aeronautical literature, as I discuss in the next section.

\section{The art of flying}

Boltzmann had a keen interest in Aeronautics and followed closely many pioneers in their attempts to
build a flying machine, particularly Otto Lilienthal in Berlin and
Wilhelm Kress in Vienna. The origins of this fascination of his are hard to trace. It is probably a natural consequence of his involvement with still unresolved theoretical questions of Aerodynamics, particularly the
problem of viscosity friction, for one should not forget that this was one of the main problems statistical
mechanicists were trying to tackle.  His interest might have also been stirred by
the sheer challenges the practical problem of flying presented, as he puts at the beginning of his lecture~\cite{boltzmannluft}:
\begin{quote}
 ``\textit{The number of unsuccesful projects in this field [Aeronautics] is legion.
However, throughout the ages, from the legendary Daedalus to Leonardo da Vinci, the greatest minds
have delved into this problem. There is no other challenge more attractive to Mankind than this one...what about Man, whose trains can overrun the fastest racing horse and whose ships, notwithstanding
their gigantic dimensions, can manouver with such ease and swiftness so as to mock the fish in their
art? Won't He ever be able to follow the birds in the skies?}''
\end{quote}

His role in the development of aeronautics was twofold: on the one hand he publicly championed the cause for more funding and governmental support in this new area of practical research, first in a lecture and
later in a newspaper article~\cite{hoeflechner1994i}. On the other hand he had a more indirect role: he is
not known to have written any theoretical paper on the topic but, as we may conclude, he stimulated one of his assistants to do so, thus taking part in the controversy around Newton's sine-square law (see below for details).

In 1894 the \textit{Gesellschaft Deutscher Naturforscher und \"Arzte}, then one of the most prestigious scientific societies, convened in Vienna for its 66th annual meeting. Boltzmann was invited to deliver a lecture. Fully aware of his own stature in the academic \textit{millieu} as well as the impact of such a gathering might have for those responsible for scientific policy, he chose to talk about the most recent experiments of Lilienthal, Kress and also Hiram S. Maxim, an american engineer living in England. His talk is a review of the most recent field tests with
heavier-than-air flying machines -- aerostats had been around for quite a while, but Boltzmann dismisses them as unpractical (in terms of mass transportation), for these were still marred by a series of disavantages~\cite{boltzmannluft}:
\begin{quote}
``\textit{If one wants a balloon to lift a man to the skies it would need to have a volume about a thousand times greater indeed...however the deployment of such huge bodies runs counter to their principal characteristic, that is light manouverability. If one wants to use a balloon one has to give up speed.}''
\end{quote}
From this he concluded that airplanes were the only reasonable solution to air transportation:
\begin{quote}
``\textit{This was just the first step towards the invention of a manouverable airship. However, that one
may also use the high speeds needed to overcome winds as a means of carrying loads can be observed
in the flight of birds of prey which, after reaching high velocities, move through the air almost without beating
their wings. So we are led to flying machines which do not depend on the lift of some gas lighter than
air, but which use the kinetic energy of a mechanism to carry loads through the skies.}''
\end{quote}
At this point he makes his heartfelt praise of Lilienthal's pioneering studies with gliders, the man who set the standards for all future studies in the area. However, true to his scientific spirit, he calls
his audience's attention to what he thought would be
the right propulsion mechanism an airplane would need: propellers. Lilienthal believed that flying machines
would achieve their goal by imitating birds in their art, \textit{i.e.} by the beating of wings. He conducted extensive studies on the profile of birds' wings (particularly storks) and designed his gliders' wings accordingly. This way he came close to the profiles we use nowadays and was the first to
realize the importance of a curved wing. The title of his book ``\textit{The flight of birds as foundation for the art of flying}'' leaves little room as to what his beliefs really were~\cite{lilienthal89}. At this point Boltzmann disagrees however, as he says~\cite{boltzmannluft}:
\begin{quote}
``\textit{Lift is caused by air resistance acting on a curved surface, a principle Wellner~\footnote{Georg Wellner (1846-1909), Czech engineer of austrian origin, professor at the University of
B$\check{r}$no (Br\"unn).} and Lilienthal most precisely measured...however the necessary horizontal speed
of an airplane can be attained by some sort of beating of wings, in which case it closely resembles a
bird, or by means of an airscrew... .}''
\end{quote}
Notwithstanding the high steem he had for Lilienthal, he could not fail to point out that
\begin{quote}
``\textit{According to Lilienthal the airplane must be divided into two halves, which move as the wings
of a bird when flying. This way one may avoid the so-called slip of the propeller and consequently the power loss due to creation of eddies, as Lilienthal believes. This
point alone I already doubt, since through the beating of the wings much of the work done when pushing them
down gets lost when pulling them up again, while with an airscrew one optimally employs the advantageous
principle of the inclined plane. Maxim's propellers actually work with very little slip. Moreover,
by dividing the airplane into movable parts one impairs its simplicity and strength. The beating of
wings cannot be attained without serious complications and a considerable friction of its parts; it
does not run smoothly and is not finely tunable as a propeller. The [theoretical] calculations are also
much more complicated.}''
\end{quote}

At this point he calls attention upon the little known Wilhelm Kress, his countryman. Kress studies suffered from insufficient funding but in spite of all difficulties he was among the first to test
surface controls on wings (flaps and ailerons) and the use of propellers. Boltzmann used all his rethorics and a great deal of patriotism to impress his
audience on what their contrymen could achieve~\cite{boltzmannluft}: 
\begin{quote}
``\textit{..how much have Germans achieved with sparse means but [only through] the acuteness of their
minds! Who dares question this, here in Vienna, the place where the Magic Flute, the Missa Solemnis and the Ninth Symphony were composed? Let the whole wide world try imitating us, if they can!}"
\end{quote}
Rather then showing an ingrained chauvinism of Boltzmann, this statement seems more of a rethorical
technique to captivate his audience. As he said, they should do something for it they did not the British
would beat them in this race:

\begin{quote}
``\textit{...I contemplate proposing to the finance committe of our newly founded Natural Society, which is
actually still sort of dangling in the air, to do something for air travel with their first available funds and, if money is not enough, to convince the government to do so.}''
\end{quote}

To make his cause more impressive Boltzmann brought along a small prototype of an airplane built by Kress, whose design, he emphasized, was already 14-years old. He made the prototype fly across the room, to the great amusement of his audience. According to a witness, the plane flew across the large auditorium, ``{\it landing in the arms of a lady}"~\cite{hoeflechner1994c}.

Boltzmann had, like few men of his time, travelled extensively through the european continent and
visited the United States twice. He also followed his son Arthur on a cruise through the Mediterranean,
while this was working on his PhD thesis. This certainly made him aware of the advantages air travel
would bring~\cite{boltzmannluft}:
\begin{quote}
``\textit{It is undeniable how immense the jump in transportation would be if one had manouverable airships,
compared to which railroads and ships would not even come close}.''
\end{quote}

From a theoretical point of view Boltzmann also had a few comments to make. It was believed that a heavier-than-air machine could not fly, an argument most theoreticians traced all the way back to
Newton. In Book II of his \textit{Principia} Newton dealt
with the resistance bodies experienced when moving through a fluid~\cite{newton}. For Newton it was
important to find an argument against the cartesian view that the Universe was filled with matter
through which forces were transmitted. If true, this would imply that planets should experience a resistance to their movement around the sun. Newton's motives were thus of a more fundamental character, not a practical one. In proposition $33$ he asserts that bodies moving through a fluid ``\textit{undergo a
resistance in a ratio compounded by the ratio of the square of their velocities, the ratio of the
square of their diameters and a simple ratio of their densities}.'' In modern parlance this means
that the aerodynamic force a body moving through air would experience is~\footnote{In Aeronautics one
speaks of three forces acting on a plane: its weight, a traction provided by the engines and an
aerodynamic force. The aerodynamic force is usually decomposed into two other forces: lift $L$ and drag $D$,
which act perpendicular and parallel to the direction of flight respectively.The modern relations for drag and lift closely resemble Newton's formula: $D=\frac{1}{2}c_{\scriptscriptstyle D}\rho A v^{2}$
and $L=\frac{1}{2}c_{\scriptscriptstyle L}\rho A v^{2}$. The important drag and lift coefficients $c_{\scriptscriptstyle D}$  and $c_{\scriptscriptstyle L}$ can be determined experimentally and the factor $1/2$ is introduced for convenience. The main task in Aeronautics is finding both coefficients, which may depend on a series of geometric factors.}
\begin{equation}
 F_{\scriptscriptstyle A}\propto \rho\; A v^{2}\;.
\end{equation}
were $\rho$ is the density, $A$ is the cross-sectional area and $v$ the velocity.
By the end of the 19th Century this formula was theoretically and experimentally well justified.
Based on a model of a fluid as a collection of particles, Newton was able to determine $F_{\scriptscriptstyle A}$ (proposition $34$) in the form:
\begin{equation}
\label{eq:sinesquared}
F_{\scriptscriptstyle A}=\rho A v^{2}\sin^2\theta\;.
\end{equation}
This is the famous \textit{sine-square law} where $\theta$ is Newton's `angle
of incidence' (nowadays called the angle of attack). Adapting it to a wing in the form of a flat plate with an angle of attack $\theta$ one can find the lift and drag  components of 
$F_{\scriptscriptstyle A}$
through the relations:
\begin{eqnarray}
D &=& F_{\scriptscriptstyle A}\sin\theta=\rho A v^{2}\sin^3\theta\nonumber\\
L &=& F_{\scriptscriptstyle A}\cos\theta = \rho A v^{2}\sin^2\theta\cos\theta\;,
\end{eqnarray}

These equations are the reason why so many physicists were very pessimistic about the feasibility of an airplane flying: if one were to increase lift,
it would be necessary to increase the angle of attack. But this would in turn lead to a much faster increase in drag (or, which is the same, a rapid decrease in the ratio $L/D=cotan\;\theta$). In other words the drag would become so large as to hinder flight. The predictions based on
Newton's arguments are wrong because of his treatment of a fluid as individual particles. 
In 1897 Gustav J\"ager, who had previously worked with
Stefan, became Boltzmann's assistant. Boltzmann took him under his wing and saw to it that J\"ager get a position of \textit{Extraordinarius} in Vienna. During his first year there
J\"ager published a paper entitled `On the resistance that bodies experience when moving through liquids and gases'~\footnote{\textit{Zur Frage des Widerstandes, welchen bewegten K\"orper in Fl\"ussigkeiten
und Gasen erfahren.}~\cite{hoeflechner1994ab}}. According to H\"oflechner, this work received a lot of attention also because of the
fact that Boltzmann highly praised it. The reason for Boltzmann's enthusiasm was that this paper refuted Newton and the findings of a commission headed by H. Helmholtz according to which it would be impossible for any machine heavier than air to fly~\cite{hoeflechner1994ab}. However as late as 1906, the year of Boltzmann's death, his colleague Franz Exner still expressed his doubts in an official report about the feasibility of an airplane~\cite{hoeflechner1994c}. 

An interesting `effect' of Boltzmann's dealings
with Aeronautics was that his son Arthur became a balloonist and later wrote an article on the resistance
of curved surfaces moving through the air~\cite{arthur}. This work deals mainly with experimental
results on the pressure distribution on curved surfaces in the presence of an air flow.

\section{Boltzmann's correspondence}

Boltzmann correspondence with Otto Lilienthal and Wilhelm Kress can be attested by the few letters
which survived. What seems to have
motivated his correspondence with Lilienthal was his need for more detailed information on Lilienthal's
studies while preparing his lecture. Boltzmann certainly knew of his work from his many visits to Berlin and had probably read Lilienthal's influential book of 1889. Lilienthal
was trained as an engineer and after successfully getting the patent rights for a few of his inventions,
he was able to more completely dedicate himself to his aeronautical studies. What characterizes his pioneering work is the combination of detailed theoretical studies with field tests. He paciently improved his gliders step-by-step, correcting minor design flaws and testing them exhaustively. The first known letter of Lilienthal to Boltzmann in the autumn of 1894 was
an answer to a letter Boltzmann had sent to him~\cite{hoeflechner1994d}:
\begin{quote}

{\it Dear Prof. Ludwig Boltzmann, Vienna

It took me a while to reply to your kind letter since I was waiting until I could send you copies
of my most recent articles on the technique of flight. I just happened to read in the newspaper
that you already gave a talk, whose report I followed with great interest. In the meantime allow
me to send you two more articles where I discuss propellers, which you mention at the end of your
letter. The October issue of Prometheus~\footnote{This was a weekly magazine which
reported on advances in sciences and industry, as well as commerce. It was edited in Berlin
and circulated between 1890 and 1920.} will carry an article of mine with illustrations of my
experimental flight station in the vicinities of Berlin~\footnote{Lilienthal is referring to an
artificial knoll he had built in Lichterfelde, a neighborhood southwest of Berlin where today one
can find a monument to his memory. The hill was named ``\textit{Fliegeberg}'' or flight mountain.}.

The privy counsellor von Stock, in the Hotel Kahlenberg, has one of my gliders.

Yours sincerely ... Otto Lilienthal.}
\end{quote}
In the meantime Boltzmann send a letter to Lord Rayleigh, dated October 4 1894, where
he thanks him for sending information on Hiram Maxim~\cite{hoeflechner1994e}:
\begin{quote}
{\it My dear Sir,

I thank you very much for the sending of your excellent book on the theory of sound~\footnote{The book
{\it The Theory of Sound} had just been released in London. The second edition was out 2 years later.}, as well as the synopsis of the article on Mr. Maxim, both of which I got while in Munich.

With the utmost respect, truly yours ... Ludwig Boltzmann.}
\end{quote}
On November 6 1894 Lilienthal replied to another letter he got from Boltzmann. From his answer we
may conclude that Boltzmann was interested in buying one of his gliders~\cite{hoeflechner1994f}:
\begin{quote}
{\it Dear Prof. Boltzmann, Vienna

It would much please me to send you a light machine. However I need some more detailed information on your
purposes. The light engines I build are all based on the stroking principle since,
in my opinion, circular movements lead to a great loss of power. However I believe you want an engine with
a rotatory motion.

In case you want to test it for a short time, so that a few minutes suffice, I would advise you to use
an engine running on liquid carbon dioxide. It exerts a pressure of 60 atm. and small bottles provide
great power. That way you can get a great concentration of power in short times.

I thank you for the news. Yours truly ... Otto Lilienthal.} 
\end{quote}
What Lilienthal meant by `stroking' movement was a new principle he wanted to apply to a prototype he
had just built. In the winter of of 1893/1894 he devised a flying machine with movable parts. True to
the principles expounded in his book, Lilienthal believed that flight would be achieved
by imitating birds, \textit{i.e.} by the beating of wings as a means of propulsion~\cite{nitsch1991}.
Boltzmann however was certain that the best solution would be to use a propeller (the circular or
rotatory movement Lilienthal mentions), a principle defended by Kress. Notwithstanding these differences in opinion, it seems that Boltzmann really meant buying one of Lilienthal's prototypes. This is attested by a letter of November 21 where he asks Lilienthal to make a price proposal~\cite{hoeflechner1994g}. On December 10 Lilienthal finally answered~\cite{hoeflechner1994h}:
\begin{quote}
{\it Dear Prof.,

in reply to your kind letter of [November] 21 I did some quick calculations as to how a rotatory machine,
that is a machine with a rotating engine, could be assembled. I would build a three-cylinder engine,
of which only one would be directly fed with carbon dioxide. This way one would get a triple expansion
without dead point. This engine, which in a 10-cycle-per-second regime produces one HP effective, would
not weigh more than 2 kg. To have it running for 2 minutes you would need 1 kg of gas. A bottle with
such a volume weighs 3 kg plus a heater of 1 kg to keep the gas from freezing. This sums up to
7 kg for one HP during 2 minutes. For most experiments 2 minutes should be enough. For 10 minutes
it would be necessary to carry 5 kg of gas and the whole equipment would weigh about 25 kg. I estimate
the price of this machine at about 1,000 Marks.

Respectfully yours ... Otto Lilienthal.}
\end{quote}
The engine Lilienthal refers to was a normal stroke (piston) engine fueled by pressurized
CO$_2$~\footnote{According to Nitsch these bottles were common at the time
and could be bought in specialized stores~\cite{nitsch1991}.}. As it was a three-stroke engine Lilienthal proposed combining three of them to get a 
{\it triple expansion without dead point}. Boltzmann most certainly did not buy the prototype, as
Lilienthal's company records do not list him among its clients~\cite{nitsch1991}. Moreover
such an unusual purchase would not go unrecorded (the famous physicist
George FitzGerald not only had of one of Lilienthal's gliders but also tested it in Dublin. Another
famous client was Nikolai Zhukovsky, the father of russian aviation who visited Lilienthal on different
occasions).

We now come to Boltzmann's next public manifest. This was a letter sent to
the newspaper {\it Neues Wiener Tagblatt} and published on July 10 1896, where he
makes reference to his acclaimed lecture of 1894. From its content one concludes that
critiques on the feasibility of Kress's prototype surfaced~\cite{hoeflechner1994i}:
\begin{quote}
{\it I have already expressed my strong conviction during the Congress of Natural Scientists in
Vienna that, among all manouverable flying apparatuses, gliders are those which have the highest chance of becoming
successful. Maxim did build a flying machine that took off with people on board, but he did not
succeed in solving the stability problem. His machine dropped suddenly, risking the lives of all
those in it.

However a solution to this problem had already been found long ago by Mr. Kress from Vienna. I just
became aware of it immediately after my lecture. It is true that he managed to get [stability] on
a small-scale model, which he actually most successfully demonstrated during my talk. Now doubts have
arisen as to whether this stability can be attained for larger models. Against these opinions I would like to say that this must be even simpler, once the necessary resources for their
construction are made available. My opinion is confirmed by recent experiments of Langley: he built
a glider which, in spite of not being able to carry a man, was much larger than Kress's. This was
only possible because instead of using rubber bands~\footnote{The models Kress built where propelled by rubber bands properly twisted.}, he used a marvelous and light 1 1/2 HP-steam engine.
According to his own reports and that of Alexander Graham Bell, the inventor of the telephone, we must believe that Langley really got enough stability. His model does not seem much different from that of
Kress. I believe that Mr. Kress will get a stable model by means of careful reasoning. Langley already
plans to build a larger one. After these results and my talk there can be no doubt that we will
manage to build light and powerful engines as well as stable prototypes, and I believe that within
five years Man will make His first long-haul flight to a predetermined destination -- and this will certainly
take place in America in case my plea for the support of austrian inventors be so successful as it was
during the Conference.

To avoid misunderstandings, I would like to note that whatever one accomplishes, it won't be a
manouverable airplane for practical use yet. However, it will be the most fundamental and important step
toward this goal. Only when we manage flying for a longer time will improvements follow...Professor Ludwig Boltzmann.}
\end{quote}

The last document is a letter of Kress
dated August 12 1896. Kress thanks Boltzmann for this article but at the same time criticizes him for being, according to Kress, overcautious. One cannot fail to notice, at the end of the letter, a certain pleading attitute
of Kress~\cite{hoeflechner1994k}:
\begin{quote}
{\it Most steemed Privy Counsellor,

I would like to thank you immensely for the clear and most favorable article of yours regarding my endeavors [published] on the 10th of this month in the {\it Neues Wiener Tagblatt}. However the last sentence `to avoid misunderstandings' shows once again your caution and unnecessarily damps a possible growing
enthusiasm for gliders. This sentence certainly does not come from your very heart, for one notices a coldness
opposite to the passionate tone previously employed. The opponents of gliders,
who think they could do better, could hold themselves to these very last words. At the same time prospective financial supporters
could become awry, for who in these days gives his money away for purely idealistic goals? I have just
read in {\it L'Aeronaute}~\footnote{A journal published in Paris between 1868 and 1906.} that M. Kane from
Chicago is building 3 different types of gasoline engines following Denington's design:
\begin{itemize}
\item[] 1 HP weighing 13 kg.
\item[] 2 HP with 18 kg or 1 HP with  9 kg.
\item[] 4 HP with 22 kg or 1 HP with 5 kg.
\end{itemize}
The interesting point about these models is that they do not require water cooling and need only
half a liter of gasoline per hour per HP. I will ask Chanute in Chicago about it. By the way, I would
like to say once more that if I had the necessary [financial] means I would quickly acquire a light
engine to make a glider fly.

Most respectfully yours... Wilhelm Kress.}
\end{quote}
An answer to this letter is unknown. Kress does not seem to have been that fair with Boltzmann since,
as Kress attests in his own book~\cite{kress1905}, from 1895 onwards he could count on military as well as Boltzmann's own financial support.

Had Boltzmann lived to see airplanes flying, he would probably be sad to see that the words he said at the beginning of his talk in Vienna would unfortunately
become reality~\cite{boltzmannluft}:

\begin{quote}
``\textit{Our modern armies would stand as much a chance against an iron airplane, flying by unassailable, dropping dynamite from the heights, as a roman army would have against a battery of cannons.}''
\end{quote}

Right at the end of his talk Boltzmann concludes:
\begin{quote}
``\textit{Only a first-rate genius can solve the problem [of airplane manouverability]. And this
inventor, besides being a genius, must also be a hero; not without great effort will the elements to
be harnessed give up their secrets. Only those who have the courage to trust their lives to these
new elements, cunningly circumventing its caveats, will stand a chance of victory over the dragon which
hides the treasures of this discovery from mankind.}''
\end{quote}
Two years after these words Otto Lilienthal, whom Boltzmann highly praised, died as a consequence of an accident while on one of his routine flights. Unfortunately there is no records of Boltzmann on this untimely death, as his own untimely death prevented him from seeing his dream become reality.

\section{Conclusions}

As the historian of science A. Vucinich once wrote, we must be careful not to exaggerate the role
of socio-cultural conditions in stimulating scientific discoveries~\cite{vucinich}. However, technological
advances are to a great measure dictated by the needs of society and science may grow as a response
to technological challenges. According to Vucinich ``{\it ... No
competent scientist would deny the essentially utilitarian quality of
scientific knowledge; but none would go on to argue that the scientific
community should concern itself exclusively with solving the acute problems
of the day}"~\cite{vucinich}. These words could well have been said by Boltzmann himself. He had a profound interest in the technological developments of his times, particularly
aviation, an interest which in my opinion went beyound that of an educated scientist. As I argued in this essay, when the opportunity presented itself for him to intervene publicly and promote aviation he did not waver: he lectured a highly prestigious audience in 1894 and sent a letter two years later to Vienna's important
\textit{Neues Wiener Tageblatt} urging authorities to provide more funding lest America beat Austria
in the race. From the point of view of basic research he supported his assistant
Gustav J\"ager who published an article countering the accepted view that heavier-than-air machines
could not fly and probably was responsible for his son Arthur's later involvement with Aerodynamics. His ideas in Physics were revolutionary, but so was his belief in Aviation: when time came for him to openly
position himself against the majority who did not believe that flying would one day become reality, he did so without hesitation. 

\section{Acknowledgments}
I would like to thank the Alexander von Humboldt Foundation for the financial support and the TP3 Group at W\"urzburg for hospitality and discussions. Dr. B. Lukasch from the Lilienthal Museum in Anklam, Germany, kindly sent me material on O. Lilienthal and a copy of A. Boltzmann's article. All translations from the original texts are my own. This article is an extended and revised version of an essay I wrote in Portuguese during the celebration of Boltzmann's Year (2006) which was published in~\cite{dahmen1}.

\end{document}